\documentclass[11pt]{article}
\usepackage[a4paper,lmargin=1in,rmargin=1in]{geometry}
\usepackage{amsmath}
\usepackage{amsfonts}
\usepackage{amsbsy}
\usepackage{amsthm}
\usepackage{accents}
\usepackage{tensor}
\usepackage{enumitem}
\newtheorem{thrm}{Theorem}

\theoremstyle{definition}

\DeclareMathOperator{\End}{End}
\DeclareMathOperator{\id}{id}
\DeclareMathOperator{\tr}{tr}
\DeclareMathOperator{\Tr}{Tr}

\newcommand{\rmi}{\mathrm{i}}
\newcommand{\rmd}{\mathrm{d}}

\newcommand{\symped}[1]{\accentset{S}{#1}}
\newcommand{\sympman}{\mathcal{M}}
\newcommand{\bund}{\mathcal{E}}
\newcommand{\ebund}{\End(\bund)}
\newcommand{\wbund}{\mathcal{W}}
\newcommand{\connsymp}{\partial^{S}}
\newcommand{\connbund}{\partial^{\bund}}
\newcommand{\connend}{\partial^{\ebund}}
\newcommand{\gambund}{\Gamma^{\bund}}
\newcommand{\gambunddot}{\dot{\Gamma}^{\bund}}
\newcommand{\curvbund}{R^{\bund}}
\newcommand{\curvsymp}{\symped{R}}

\newcommand{\bsh}{\mathsf{h}}
\newcommand{\ricciend}{\underline{R}}
\newcommand{\riemannend}{\underline{\underline{R}}}
\newcommand{\modricciend}{\underline{\breve{R}}}
\newcommand{\modriemannend}{\underline{\underline{\breve{R}}}}
\newcommand{\tenmodrl}[1]{\tensor{\undertilde{\breve{R}}}{#1}}
\newcommand{\tent}[1]{\tensor{\Theta}{#1}}
\newcommand{\tenr}[1]{\tensor{R}{#1}}
\newcommand{\lbund}{\mathcal{L}}
\newcommand{\tenrl}[1]{\tensor{\undertilde{R}}{#1}}
\newcommand{\tenth}[1]{\tensor{\theta}{#1}}

\begin{document}
\linespread{1.3}

\title{\textsc{Involution in quantized endomorphism bundle\\ 
and reality of noncommutative gravity actions}}
\date{}
\author{Micha{\l} Dobrski\footnote{michal.dobrski@p.lodz.pl}
\\
\small
\emph{Centre of Mathematics and Physics}
\\
\small
\emph{Technical University of {\L}\'od\'z,}
\\
\small
\emph{Al.~Politechniki 11, 90-924 {\L}\'od\'z, Poland}}
\maketitle
\abstract{It is shown that for arbitrary connection in the vector bundle compatible with some Hermitian metric, the corresponding Fedosov trace functional commutes with involution generated by this metric. This result is then used to prove that certain noncommutative gravity actions are real in all powers of deformation parameter.}

\section{Introduction}
In \cite{dobrski-sw,dobrski-ncgr} the Fedosov formalism of deformation quantization of endomorphism bundle has been considered as a tool for building geometric noncommutative field theories. Roughly speaking, the main result of \cite{dobrski-sw} states that the theory of Seiberg-Witten map is (more or less) equivalent to a deformation quantization of endomorphism bundle, while in \cite{dobrski-ncgr} this observation is used to build some geometric models of noncommutative vacuum general relativity. Results obtained in \cite{dobrski-ncgr} exhibit nice characteristic: first order imaginary corrections to field equations vanish. (This is very common property of noncommutative gravity theories based on Seiberg-Witten map -- compare e.g. \cite{calmet1,calmet2,mukherjee,mukherjee2,chamseddine1,chamseddine2,chamseddine3,cardella,chaichian,garciacompean}). However, it was not clear whether higher order imaginary terms also disappear. In the following short note we investigate Fedosov $*$-product of endomorphisms originating in connection compatible with some Hermitian metric $\bsh$. We show that in such case the corresponding Fedosov trace functional commutes with $\bsh$-induced involution. This result is then used to analyze reality of actions constructed in $\cite{dobrski-ncgr}$ and it is concluded that they are purely real in all powers of deformation parameter.

\section{Hermitian metric and Fedosov quantization}

\subsection{Involution and the $*$-product}
This subsection contains collection of some well know facts (elaborated by Waldmann in \cite{waldman}; compare also \cite{bur-wald}) recalled here for the purpose of fixing some necessary background. We are not going to repeat the Fedosov construction in full detail. Instead, we just quickly follow original presentation of \cite{fedosov0,fedosov} relating it to the Hermitian metric. Notations and conventions are compatible with \cite{dobrski-sw, dobrski-ncgr} and, in turn, with \cite{fedosov0,fedosov}.

Consider a finite dimensional complex vector bundle $\bund$ over a real smooth manifold $\sympman$. Let $\bund$ be equipped with a Hermitian metric $\bsh$. It induces an involution $(\cdot)^+$ in the endomorphism bundle $\ebund$ defined by the relation
\begin{equation}
\bsh(Av,w)=\bsh(v,A^+ w),
\end{equation}
for an arbitrary section $A \in C^{\infty}(\ebund)$ and all $v,w \in C^{\infty}(\bund)$. Clearly, one can define action of $(\cdot)^+$ on complex differential forms with values in $\ebund$ by antilinear extension of the rule
\begin{equation}
(A\otimes\eta)^+=A^+ \otimes \bar{\eta},
\end{equation}
for $A \in C^{\infty}(\ebund)$ and $\eta \in C^{\infty}(\Lambda(\sympman))$.

Let $\sympman$ be the Fedosov manifold \cite{gelfand,bielgutt} endowed with a symplectic form $\omega$ and a symplectic connection $\connsymp$ with curvature tensor denoted by $\tensor{\curvsymp}{^i_{jkl}}$. Also, let $\connbund$  be a linear connection in $\bund$, compatible with the metric $\bsh$, i.e. $\connbund_i \bsh(v,w)=\bsh(\connbund_i v, w) + \bsh(v, \connbund_i w)$. Connection $\connbund$ induces connection $\connend$ in $\ebund$, for which the relation
\begin{equation}
\label{cendcommuteswithinv}
(\connend A)^+ = \connend (A^+ ), 
\end{equation} 
holds for arbitrary $A \in C^{\infty}(\ebund)$. It follows that the curvature $2$-form  \mbox{$1/2 \curvbund_{ij}\rmd x^i \wedge \rmd x^j$} of $\connbund$ is anti-self-adjoint
\begin{equation}
\label{curvaherm}
{\curvbund_{ij}}^+ = - \curvbund_{ij}.
\end{equation}

Introducing over $\sympman$ the Weyl algebras bundle with coefficients in $\ebund$ (denoted by $\wbund$), one observes that $(\cdot)^+$ can be extended to $\wbund \otimes \Lambda$ by\footnote{Here $a_{i_1 \dots i_p j_1 \dots j_q}(x)$ are components of some $\ebund$-valued covariant tensor field at $x \in \sympman$, and $y \in T_x\sympman$}
\begin{equation}
a^+(x,y)=\sum_{k,p \geq 0} h^k a^+_{i_1 \dots i_p j_1 \dots j_q}(x)y^{i_1} \dots y^{i_p} \rmd x^{j_1} \wedge \dots \wedge \rmd x^{j_q}.
\end{equation}
It can be easily shown that for the fiberwise Moyal product $\circ$ the formula
\begin{equation}
\label{circinv}
(a \circ b)^+ = (-1)^{rs} b^+ \circ a^+
\end{equation}
holds, with $a \in \wbund \otimes \Lambda^r$, $b \in \wbund \otimes \Lambda^s$. Consequently, for commutators $[a,b]=a \circ b - (-1)^{rs} b \circ a$ above relation induces
\begin{equation}
\label{comminv}
[a,b]^+ = -[a^+,b^+].
\end{equation}
Connections $\connsymp$ and $\connend$ give rise to the connection $\partial$ in $\wbund \otimes \Lambda$. From (\ref{cendcommuteswithinv}) and reality of $\connsymp$ one infers that
\begin{equation}
\label{wbundconninv}
(\partial a)^+ = \partial (a^+ ), 
\end{equation} 
for arbitrary section $a$ of $\wbund \otimes \Lambda$. Then, it is straightforward that operators $\delta a =\rmd x^k \wedge \frac{\partial a}{\partial y^k}$ and 
$\delta^{-1}a_{km}=\frac{1}{k+m}y^s \iota \left(\frac{\partial}{\partial x^s}\right) a_{km}$ (for $a_{km}$ with $k$-fold $y$ and $m$-fold $\rmd x$) commute with the involution
\begin{equation}
\label{deltasinv}
(\delta a)^+ = \delta (a^+) \quad \textrm{ and } \quad (\delta^{-1} a)^+ = \delta^{-1} (a^+).
\end{equation}

Now, the Abelian connection is determined by a $1$-form $r$ defined as the solution of
\begin{equation}
\label{fedo_abeliter}
r=r_0 + \delta^{-1}\left( \partial r + \frac{\rmi}{h}r \circ r\right)
\end{equation}
for $r_0=\delta^{-1}R$, where\footnote{The indices are manipulated by means of symplectic form, i.e.  $\curvsymp_{ijkl}=\omega_{is}\tensor{\curvsymp}{^s_{jkl}}$.} $R=1/4\, \curvsymp_{ijkl}y^i y^j \rmd x^k \wedge \rmd x^l- \frac{\rmi h}{2}\ \curvbund_{kl} \rmd x^k \wedge \rmd x^l$. Applying involution to both sides of (\ref{fedo_abeliter}) and using relations (\ref{curvaherm}), (\ref{circinv}), (\ref{wbundconninv}), (\ref{deltasinv}) one gets 
\begin{equation}
r^+=r
\end{equation}
from the uniqueness of the solution of (\ref{fedo_abeliter}).
Hence, the Abelian connection $D=-\delta+\partial+\frac{\rmi}{h}[r,\cdot\,]$ commutes with the involution
\begin{equation}
\label{abelianinv}
(Da)^+=D(a^+).
\end{equation}
In turn, the same stays true for the ``quantization lifting'' $Q$ defied as the solution of
\begin{equation}
\label{fedo_q_iter}
b=a + \delta^{-1}(D+\delta)b
\end{equation}
with respect to $b$, and for its inverse $Q^{-1}=\id-\delta^{-1}(D+\delta)$, i.e.
\begin{equation}
\label{qqinv}
Q(a)^+ = Q(a^+) \quad \textrm{ and } \quad Q^{-1}(a)^+ = Q^{-1} (a^+).
\end{equation}
(Recall, that $Q$ maps $C^{\infty}(\ebund)[[h]]$ to $\wbund_D$, where $\wbund_D$ is subalgebra of algebra of sections of $\wbund$, defined by the requirement of flatness $Da=0$).
As a final conclusion we observe that $(\cdot)^{+}$ is an involution with respect to the Fedosov product of endomorphisms given by $A*B:=Q^{-1}(Q(A) \circ Q(B))$. Indeed, equation (\ref{qqinv}) together with (\ref{circinv}) yield
\begin{equation}
\label{hermstprod}
(A*B)^+=B^+*A^+ ,
\end{equation}
while the remaining requirements are fulfilled in the straightforward way.

\subsection{Involution and the trace functional}
Now, we are going to prove the following theorem
\begin{thrm}
\label{my_thrm}
For arbitrary $\bsh$-compatible connection $\connbund$, the property
\begin{equation}
\label{inv_trace}
\tr_*(A^+)=\overline{\tr_*(A)}
\end{equation}
holds for the corresponding Fedosov trace functional. 
\end{thrm}
Such relation is quite natural, hence one can expect that it is well-known for many (or at least for  someone). However, the author was unable to point out any reference clearly stating (\ref{inv_trace}). Thus, let us analyze the proof in some detail.

First, let us concern the behavior of involution under some $*$-product isomorphism. Let ${\connbund}^{(t)}$ and ${\connsymp}^{(t)}$ be (possibly local) homotopies of $\bsh$-compatible and symplectic connections\footnote{Thus, we demand ${\connbund}^{(t)}$ and ${\connsymp}^{(t)}$ to be, respectively, $\bsh$-compatible and symplectic for each $t$.} such that ${\connbund}^{(0)}=\connbund$, ${\connsymp}^{(0)}=\connsymp$ and both ${\connbund}^{(1)}$, ${\connsymp}^{(1)}$ are flat. They generate family of products $*_{t}$ described by the homotopy of Abelian connections $D_t=-\delta+{\connbund}^{(t)}+[r(t),\cdot\,]$, where $r(t)$ is obtained due to formula (\ref{fedo_abeliter}).
For $t=0$ the initial product is recovered (i.e. $*_0=*$), and for $t=1$ one is dealing with a trivial algebra described by the Moyal formula, for which the partial derivatives are replaced by flat covariant ones. We are going to use symbol $*_T$ instead of $*_1$ in such case.
Obviously, the relations (\ref{cendcommuteswithinv}), (\ref{curvaherm}), (\ref{wbundconninv}), (\ref{abelianinv}), (\ref{qqinv}) and (\ref{hermstprod}) hold for each~$t$. 

The isomorphism between $\wbund_{D_0}$ and $\wbund_{D_t}$ can be constructed in the following way (\cite{fedosov} section 5.3). Define 
\begin{equation}
\gamma(t)=1/2 \Gamma_{ijk}(t) y^{i}y^{j}\rmd x^k-\rmi h \gambund(t)+r(t),
\end{equation} 
where $\gambund(t)$ comes from ${\connbund}^{(t)}=\rmd+\gambund(t)$ and $\tensor{\Gamma}{^i_{jk}}(t)$ denotes connection coefficients of ${\connsymp}^{(t)}$. Observe that the derivative $\dot{\gamma}$ is well defined $1$-form, independent of the particular choice of the frame in $\bund$ and local coordinates on $\sympman$. Introduce Hamiltonian $H(t)$ as a solution of $D_t H=\dot{\gamma}$. It can be chosen as $H(t)=-Q_t \delta^{-1}\dot{\gamma}(t)$ yielding 
\begin{equation}
H(t)=Q_t \bigg(\rmi h \gambunddot_j (t)y^j-\frac{1}{6} \dot{\Gamma}_{ijk}(t) y^i y^j y^k\bigg).
\end{equation} 
A quick calculation ensures that from $\bsh$-compatibility of ${\connbund}^{(t)}$ it follows that ${\gambunddot_i}^+=-\gambunddot_i$, and hence, by (\ref{qqinv}), the Hamiltonian is self-adjoint
\begin{equation}
\label{hamselfad}
H^+ (t)=H(t).
\end{equation}
Then, the family of isomorphisms $T_t:\wbund_{D_0} \to \wbund_{D_t}$ is given by the unique solution of the Heisenberg equation
\begin{equation}
\label{triv_heis}
\frac{\rmd a}{\rmd t}+\frac{\rmi}{h}[H(t),a]=0,
\end{equation}
i.e. $T_t(a(0))=a(t)$. At the level of sections of $C^{\infty}(\ebund)[[h]]$ it corresponds to the family of isomorphisms transporting $*_0$ to $*_t$. That is for $M_t(A)=Q^{-1}_t(T_t(Q_0(A)))$ the the property $M_t(A *_0 B) = M_t(A) *_t M(B)$ holds. For $t=1$ we are going to omit the subscript and to write
\begin{equation}
\label{M_iso}
M(A * B) = M(A) *_T M(B).
\end{equation}
Applying the involution to both sides of (\ref{triv_heis}) and using (\ref{comminv}), (\ref{hamselfad}) we arrive at $T_t(a^+)=T_t(a)^+$ by the uniqueness of the solution of (\ref{triv_heis}). Consequently
\begin{equation}
\label{M_invol}
M(A^+)=M(A)^+ .
\end{equation}

We are almost ready to prove the relation (\ref{inv_trace}). Recall that for the compactly supported sections of $\ebund$ the Fedosov trace functional is uniquely defined (up to a normalizing constant) by the property
\begin{equation}
\tr_* (A*B)=\tr_* (B*A)
\end{equation}
and the requirement of invariance under $*$-product isomorphisms, i.e.
\begin{equation}
\label{trisoinvar}
\tr_{*_1}(A)=\tr_{*_2}(M(A)),
\end{equation}
for $M=\id+O(h)$ being an isomorphism $M(A *_1 B)=M(A) *_2 M(B)$. The fully explicit formula for $\tr_*$ is not known in the general case, but for trivial algebras the trace is given by the integral\footnote{In order to keep the notations compatible with \cite{dobrski-ncgr}, we make noncanonical (compare \cite{fedosov}) choice of the normalizing constant by setting it to $1$.}
\begin{equation}
\tr_{*_T}A=\int \Tr A \frac{\omega^n}{n!},
\end{equation}
where $\Tr$ stands for the trace of matrix. For $\tr_{*_T}$ the relation 
\begin{equation}
\label{inv_triv_trace}
\tr_{*_T}(A^+)=\overline{\tr_{*_T}(A)}.
\end{equation}
holds trivially.

In the original presentation of \cite{fedosov} the existence and uniqueness of $\tr_*$ have been proven by taking appropriate set of trivializations to Moyal algebra and a compatible partition of unity. Let us make use of some variant of this procedure. Let $\{\mathcal{O}_i\}$ be a covering of the support of some section $A \in C^{\infty}(\ebund)[[h]]$, such that for each $\mathcal{O}_i$ one can construct $\bsh$-compatible and symplectic homotopies flattening $\connbund$ and $\connsymp$ respectively, and in turn obtain isomorphism $M^i$ of type (\ref{M_iso}). (By compactness we can always make $\{\mathcal{O}_i\}$ to be finite). Also, let $\{ \rho_i \}$ be a compatible real partition of unity. With these data and in virtue of  (\ref{hermstprod}), (\ref{trisoinvar}), (\ref{M_invol}), (\ref{inv_triv_trace}), we get
\begin{multline}
\tr_* (A^+)= \sum_i \tr_* (\rho_i * A^+)=\sum_i\tr_* ((A * \rho_i)^+)=\sum_i \tr_{*_T} (M^i(A*\rho_i)^+)\\
=\overline{\sum_i \tr_{*_T} (M^i(A*\rho_i))}
=\overline{\tr_*(A)},
\end{multline}
and complete the proof of theorem \ref{my_thrm}.

\section{Reality of noncommutative relativity actions}
In \cite{dobrski-ncgr} several models of geometric noncommutative gravity were proposed. The construction was carried out by means of deformation quantization of endomorphism bundle 
and formulated on arbitrary Fedosov manifold $(\sympman,\omega,\connsymp)$. The considered actions read as follows. (The notation $\breve{A}=vA$ is used in further considerations. Here $v : \sympman \to \mathbb{R}$ is a function defined by proportionality between metric and symplectic volume forms i.e. $\sqrt{-g} \rmd x^1\wedge\dots\wedge \rmd x^{2n} = v \frac{\omega^n}{n!}$). 
\begin{itemize}
\item
\begin{equation}
\widehat{\mathcal{S}}_{EH_{1A}}=\tr_{*_{EH_1}}(\modricciend),
\end{equation}
where the dynamical variable is metric $g_{ab}$ in $T\sympman$. The $*_{EH_1}$ is the star product in $C^{\infty}(\End(T \sympman))[[h]]$ generated by the metric Levi-Civita connection $\partial^{T \sympman}$. $\ricciend$ stands for the endomorphism of $T \sympman$ given by $\tensor{\ricciend}{^a_b}=\tensor{R}{^a_b}=g^{ac} R_{cb}$, where $R_{cb}$ is Ricci tensor of $\partial^{T \sympman}$. 
\item
\begin{equation}
\widehat{\mathcal{S}}_{EH_{1B}}=\tr_{*_{EH_1}}(\ricciend *_{EH_1} V)
\end{equation}
Here we use the same star product structure as in $\widehat{\mathcal{S}}_{EH_{1A}}$, but we force correct volume form at the undeformed level by endomorphism $V=\breve{1}=v 1$.
\item
\begin{equation}
\widehat{\mathcal{S}}_{EH_{2A}}=\tr_{*_{EH_2}}(\modriemannend)
\end{equation}
This time $*_{EH_2}$ is the star product in $C^{\infty}(\End(T \sympman \otimes T \sympman))[[h]]$ generated by the connection $\partial^{T \sympman \otimes T \sympman} = \partial^{T \sympman} \otimes 1 + 1 \otimes \partial^{T \sympman}$, where $\partial^{T \sympman}$ is again Levi-Civita connection corresponding to the dynamical variable $g_{ab}$. $\riemannend$ denotes the endomorphism of $T \sympman \otimes T \sympman$ defined by $\tensor{\riemannend}{^{ab}_{cd
}}=\tensor{R}{^{ab}_{cd}}=g^{bs}\tensor{R}{^{a}_{scd}}$, with $\tensor{R}{^{a}_{bcd}}$ being the Riemann tensor of metric connection $\partial^{T \sympman}$. 
\item
\begin{equation}
\widehat{\mathcal{S}}_{EH_{2B}}=\tr_{*_{EH_2}}(\riemannend *_{EH_2} V)
\end{equation}
Here one deals with the variant of $\widehat{\mathcal{S}}_{EH_{2A}}$ with endomorphism $V=\breve{1}=v 1$ used.
\item
\begin{equation}
\widehat{\mathcal{S}}_{P}=\tr_{*_P}(\tenmodrl{}*_P\tent{})
\end{equation}
Consider the vector bundle $\lbund$ over $\sympman$, for which  $SO(3,1)$ transformations preserve the canonical form of the Lorentzian metric $\eta_{AB}$. The bundle $\lbund$ is equipped with some metric-compatible connection $\partial^{\lbund}$. 
The corresponding curvature is given by $\tenrl{^A_{Bij}}$. The orthonormal tetrad field $\tenth{^A_{b}}$ induces the metric $g_{ab}=\tenth{^A_{a}}\eta_{AB}\tenth{^B_{b}}$ and the metric connection $\nabla$ in $T\sympman$ (not necessarily torsionless). 
The star product $*_P$ is a multiplication in $C^{\infty}(\End(\lbund \otimes T\sympman))[[h]]$ generated by the connection in $\lbund \otimes T\sympman$ taken as $\partial^{\lbund \otimes T\sympman}=\partial^{\lbund} \otimes 1+1 \otimes \nabla$. The following two endomorphisms of $\lbund \otimes T\sympman$ are used: $\tenrl{^A_B^{a}_{b}}$ (defined by the curvature of $\partial^{\lbund}$ and the tetrad which raises index ${}^{a}$), and $\tent{^A_B^{a}_{b}}=\tenth{^{Aa}}\tenth{_{Bb}}$. The dynamical variables are the tetrad field $\tenth{^A_{b}}$ and connection coefficients of $\partial^{\lbund}$.
\end{itemize}
Thus $\widehat{\mathcal{S}}_{EH_{1A}}$, $\widehat{\mathcal{S}}_{EH_{1B}}$, $\widehat{\mathcal{S}}_{EH_{2A}}$ and $\widehat{\mathcal{S}}_{EH_{2B}}$ correspond to deformations of Einstein-Hilbert action, while $\widehat{\mathcal{S}}_{P}$ is related to Palatini one.

Using results of previous section one can easily observe that all above action functionals are real. First, notice that in each case the star product structure is taken with respect to a connection compatible with some metric. Specifically one is dealing with
\begin{itemize}
\item metric $g_{ab}$ and its Levi-Civita connection $\nabla$ in $T \sympman$ for the case of $\widehat{\mathcal{S}}_{EH_{1A}}$ and $\widehat{\mathcal{S}}_{EH_{1B}}$,
\item metric $\tilde{g}$ in $T \sympman \otimes T \sympman$ (defined as $\tilde{g}(X,Y)=g_{ac} g_{bd} X^{ab} Y^{cd}$ for $X,Y \in T\sympman \otimes T\sympman$) and compatible connection $\partial^{T \sympman \otimes T \sympman}$ for $\widehat{\mathcal{S}}_{EH_{2A}}$ and $\widehat{\mathcal{S}}_{EH_{2B}}$,
\item metric $\tilde{\tilde{g}}$ in $\lbund \otimes T \sympman$ (given by $\tilde{\tilde{g}}(L,N)=\eta_{AB} g_{ab}  L^{Aa} N^{Bb}$ for $L,N \in \lbund \otimes T\sympman$) and metric connection $\partial^{\lbund \otimes T \sympman}$ in the case of $\widehat{\mathcal{S}}_{P}$.
\end{itemize}
Now, by switching to Fedosov deformation quantization of $\ebund$, one silently replaces all underlying bundles $\bund$ (i.e. $T\sympman$, $T\sympman \otimes T\sympman$ and $\lbund \otimes T \sympman$) by their complexifications. Under this procedure $g$, $\tilde{g}$, $\tilde{\tilde{g}}$ are naturally promoted to Hermitian metrics $\bsh$, $\tilde{\bsh}$, $\tilde{\tilde{\bsh}}$ according to the rule $\bsh(X,Y)=g(X,\overline{Y})$ (and analogously for $\tilde{\bsh}$ and $\tilde{\tilde{\bsh}}$). Metric connections become compatible with these Hermitian forms. Thus, for all star products $*_{EH_{1}}$, $*_{EH_{2}}$ and $*_{P}$, the formulae (\ref{hermstprod}), (\ref{inv_trace}) hold with involutions (call them $(\cdot)^+$, $(\cdot)^{\tilde{+}}$ and $(\cdot)^{\tilde{\tilde{+}}}$ respectively) defined by corresponding Hermitian form. Finally, one can quickly calculate that from symmetries $R_{ab}=R_{ba}$, $R_{abcd}=R_{cdab}$, $\tenrl{_{ABab}}=\tenrl{_{BAba}}$ (and reality of $v$, $\tenr{^{a}_{bcd}}$, $\tenrl{^{A}_{Bcd}}$, $\tenth{^A_b}$) it follows
\begin{gather*}
\modricciend^+=\modricciend \quad V^+=V, \quad \ricciend^+=\ricciend,\\
\modriemannend^{\tilde{+}}=\modriemannend, \quad \riemannend^{\tilde{+}}=\riemannend, \quad V^{\tilde{+}}=V,\\
 \tenrl{}^{\tilde{\tilde{+}}}=\tenmodrl{}, \quad \tent{}^{\tilde{\tilde{+}}}= \tent{}.
\end{gather*}
Finally, using (\ref{hermstprod}) and (\ref{inv_trace}) one infers that all investigated actions are real up to arbitrary power of deformation parameter
\begin{gather*}
\overline{\widehat{\mathcal{S}}\,}_{EH_{1A}}=\widehat{\mathcal{S}}_{EH_{1A}}, \quad \overline{\widehat{\mathcal{S}}\,}_{EH_{1B}}=\widehat{\mathcal{S}}_{EH_{1B}}, \\ 
\overline{\widehat{\mathcal{S}}\,}_{EH_{2A}}=\widehat{\mathcal{S}}_{EH_{2A}}, \quad  \overline{\widehat{\mathcal{S}}\,}_{EH_{2B}}=\widehat{\mathcal{S}}_{EH_{2B}}, \\ 
\overline{\widehat{\mathcal{S}}\,}_{P}=\widehat{\mathcal{S}}_{P}.
\end{gather*}
We conclude that all theories considered in \cite{dobrski-ncgr} are based on purely real actions and thus produce real field equations and real corrections to metrics.

\section*{Acknowledgments}
I would like to thank professor Piotr Kosi\'{n}ski for suggesting that ``involution can help''. Also, I am grateful to professor Maciej Przanowski for reviewing an initial version of the manuscript and helpful remarks.

\end{document}